\documentclass[showpacs,aps,pre,twocolumn, floatfix,superscriptaddress, graphics]{revtex4-1}
\usepackage[ocgcolorlinks,colorlinks=true,linkcolor=blue,citecolor=orange,linktocpage=true]{hyperref}
\usepackage{amsmath,amssymb,graphicx,color,braket,hyphenat,makeidx,xcolor,enumitem}

\newcommand{\tr}{{\rm Tr}}

\newcommand{\blue}{}

\begin{document}

\title{The squeezed thermal reservoir as a generalized equilibrium reservoir}
\author{Gonzalo Manzano}
\affiliation{Scuola Normale Superiore, Piazza dei Cavalieri 7, I-56126, Pisa, Italy}
\affiliation{International Center for Theoretical Physics, Strada Costiera 11, Trieste 34151, Italy}
\date{\today}

\begin{abstract}
We explore the perspective of considering the squeezed thermal reservoir as an equilibrium reservoir in a generalized Gibbs ensemble with two non-commuting conserved quantities. 
We outline the main properties of such a reservoir in terms of the exchange of energy, both heat and work, and entropy, giving some key examples to clarify its physical interpretation.
This allows for a correct and insightful interpretation of all thermodynamical features of the squeezed thermal reservoir, as well as other similar non-thermal reservoirs, including the 
characterization of reversibility and the first and second laws of thermodynamics.

\end{abstract}

\maketitle

\section{Introduction}

Thermodynamics is arguably one of the most robust and successful theories in physics. The source of its strength and wide scope resides on the generality and simplicity of their 
foundational principles, the laws of thermodynamics. Still nowadays they continue to provide us with new insights in a broad range of physical systems, from black holes down to the microscopic world, and entering the quantum realm \cite{Qtermo1,Qtermo2}.
In particular, one of the main questions which has recently attracted great attention is the development of a thermodynamic description of quantum effects \cite{Qeffects} and the elucidation of its role from an operational 
point of view \cite{Operational}. 

In this context, ongoing discussions concern the correct interpretation of the energetics and entropy dynamics of systems interacting with quantum non-thermal reservoirs \cite{Abah14, Man16, Nie16, Nie18}. 
These are reservoirs for which the presence of a quantum property such as coherence \cite{Scully03}, quantum correlations \cite{Lutz09} or squeezing \cite{Huang12, Lutz14, Correa14} renders its state non-thermal, that is, which cannot longer be described by a Gibbs state. 
The use of non-thermal reservoirs in heat engine setups lead to striking situations. In particular, for the case of the squeezed thermal reservoir, these situations include the possibility of work extraction from a single reservoir \cite{Man16}, 
the emergence of multiple operational regimes \cite{Man16, Nie16}, and the surpassing of Carnot's bound \cite{Huang12, Lutz14, Correa14, Abah14, Man16, Segal17}. Some of these predictions has been indeed recently demonstrated in the laboratory in a proof-of-principle 
experiment \cite{Klaers}.

Nevertheless, even if it is clear that traditional thermodynamic inequalities developed for thermal reservoirs cannot be directly applied to non-thermal ones (e.g. Carnot bound), the laws governing the thermodynamic behavior of such reservoirs still requires clarification. 
Until now, non-thermal reservoirs such as the squeezed thermal reservoir, has been essentially conceptualized as stationary \emph{non-equilibrium} reservoirs, constructed e.g. from reservoir engineering techniques \cite{Abah14, Lutz14}. Under this view, their role is similar in spirit 
as the one played by the nonequilibrium environmental conditions in flashing ratchets \cite{flashing}, molecular motors \cite{Sei11}, or information reservoirs \cite{Mandal12, Deffner13, Barato14}. This perspective raised the belief that reversible operations are forbidden when using 
non-thermal reservoirs, as explicitly manifested in most recent works focusing on the energetics of quantum engines from the point of view of ergotropy \cite{Nie18}. 

Here we show that the squeezed thermal reservoir, contrary to the previous belief, can be regarded as a generalized {\it equilibrium} reservoir \cite{Jay57}, which exchanges two non-commuting (conserved) quantities \cite{Gur16, Hal16, Los17} with the system to which it is coupled: energy and second order coherence, which 
we call asymmetry \cite{Man16}. Squeezing is a paradigmatic quantum effect rooted in Heisenberg's uncertainty principle. It can be defined as the reduction in the uncertainty of some observable at expenses of the increase in the conjugate one \cite{Ficek}, finding notable applications in quantum metrology, computation, 
cryptography and imaging \cite{Polzik}. Our proposal allows for a correct and insightful interpretation of all thermodynamical features of the squeezed thermal reservoir, as well as other similar non-thermal reservoirs, including the first and second laws of thermodynamics. 
Starting with the analysis of entropy changes in the reservoir \cite{QFT}, we are able to distinguish a work contribution associated to the transfer of squeezing, much in the spirit of chemical work for heat and particle reservoirs, or as in other work producing reservoirs \cite{Hor16}. 
Finally, we provide a generic reversible protocol for processes in contact with a squeezed thermal reservoir, and apply the framework to the experimentally relevant situation of extracting work from a single reservoir \cite{Klaers}. 
Our results show the deep link between the extractable mechanical work, coherence, and the work performed by the reservoir.


This paper is organized as follows. In Sec. \ref{Sec2} we introduce a bosonic squeezed thermal reservoir and discuss their main properties as a generalized equilibrium reservoir. 
We follow in Sec. \ref{Sec3} by considering a driven bosonic mode weakly interacting with the squeezed thermal reservoir. Here we develop a thermodynamic description based on an interaction supporting 
conservation of energy and asymmetry. In Sec. \ref{Sec4} we construct a generic reversible process allowing work and asymmetry extraction from arbitrary nonequilibrium states and apply it to the case of cyclic work extraction from a 
single reservoir in Sec. \ref{Sec5}. A specific example illustrating the difference between our approach and ergotropy-based approaches is provided in Sec. \ref{Sec6}. We conclude in Sec. \ref{Sec7} with some final remarks.

\section{The squeezed thermal reservoir} \label{Sec2}
Let us start by first defining the squeezed thermal reservoir. We consider a set of $N$ non-interacting bosonic modes with Hamiltonian $H_R = \sum_k H_R^{(k)} = \sum_k \hbar \omega_k b_k^\dagger b_k$, with bosonic ladder operators $[b_k, b_{l}^\dagger] = \delta_{k,l}$. 
Assume that each of them is in a squeezed thermal state, that is, a canonical Gibbs state at some inverse temperature $\beta_0$ to which the squeezing operator has been applied
\begin{equation} \label{eq:res}
 \rho_R^{(k)} = \mathcal{S}(\xi) \frac{e^{- \beta_0 H_R^{(k)}}}{Z_k^0} \mathcal{S}^\dagger(\xi), 
\end{equation}
where $Z_k^0 = \tr[e^{- \beta_0 H_R^{(k)}}]$ is the partition function, $\mathcal{S}(\xi) = \exp[\frac{1}{2}(b_k^2 \xi^\ast - b_k^{\dagger 2}\xi)]$ is the unitary squeezing operator and $\xi \equiv |\xi|e^{i \theta}$ is the complex squeezing parameter. 
Along this paper we will consider without loss of generality $\theta = 0$. Physically, the state in Eq. \eqref{eq:res} corresponds to the output of a parametric amplifier whose input is thermal noise \cite{Ficek, Fearn88}.

Our main motivation is to interpret the above state as a generalized Gibbs ensemble \cite{Jay57, Gur16, Hal16, Los17, Rig07, Vac11, Lan15, Perarnau16, Hor16}. 
In order to do this, we immediately notice that, since $\mathcal{S}(\xi)$ is unitary, and whenever $H_R^{(k)}$ is quadratic, we can rewrite \eqref{eq:res} in a more convenient way:
\begin{equation}\label{eq:gge}
\rho_R^{(k)} = \frac{e^{- \beta (H_R^{(k)} - \mu A_R^{(k)})}}{Z_k}.
\end{equation}
with $Z_k= \tr[e^{- \beta (H_R^{(k)} - \mu A_R^{(k)})}]$. Here we defined the following inverse temperature and a chemical-like potential of squeezing:
\begin{equation} \label{eq:chem}
 \beta \equiv \beta_0 \cosh(2\xi), ~~ \mu \equiv \tanh(2\xi),
\end{equation}
together with the quantity
\begin{equation} \label{eq:asy}
A_R^{(k)} \equiv - \frac{\hbar \omega_k}{2} \left(b_k^{\dagger 2} + b_k^2 \right),
\end{equation}
with energy units. We will refer to \eqref{eq:asy} as the second-order moments asymmetry or just the asymmetry for reservoir mode $k$, since it can be rewritten as 
$A_R^{(k)} = (\hbar \omega_k/2)(p_k^{2} - x_k^{2})$, where $[x_{k}, p_{l}] = i \hbar \delta_{k l}$ are the (dimensionless) position and momentum quadratures of the modes \cite{Man16}.
Notice that Eq. \eqref{eq:gge} follows from the property $\beta_0 \mathcal{S}(\xi) H_R^{(k)} \mathcal{S}^\dagger(\xi) = \beta (H_R^{(k)} - \mu A_R^{(k)}) + \beta_0 \sinh^2(\xi)$, 
which is a direct consequence of the squeezing operator action over ladder operators $\mathcal{S}(\xi) b_k \mathcal{S}^\dagger(\xi) = b_k \cosh(\xi) + b_k^\dagger \sinh(\xi)$.
We also notice that $\rho_R^{(k)}$ is the state maximizing the von Neumann entropy of the reservoir mode $k$, subjected to fixed averages $\langle H_R^{(k)} \rangle$ and $\langle A_R^{(k)} \rangle$ \cite{Jay57} (see Ref. \cite{Liu06} for a different proof). 
Consequently, the inverse temperature $\beta$ and the chemical-like potential of squeezing defined in Eq. \eqref{eq:chem} are the conjugate variables to the two (non-commuting) conserved quantities $H_R^{(k)}$ and $A_R^{(k)}$ respectively, 
verifying maximization (Lagrange multipliers). 

A more physical picture of the observable $A_R^{(k)}$ in Eq. \eqref{eq:asy}, can be given by considering the case of (zero-mean) Gaussian states. In such case it just corresponds to the difference between the uncertainties in 
the mode quadratures $p_k$ and $x_k$, $\langle A_R^{(k)} \rangle_{\rho_R^{(k)}} = \hbar{\omega}(\Delta p_k^2 - \Delta x_k^2)$, which is a measure of squeezing. For example, in the case of the squeezed thermal state in Eq. \eqref{eq:res} 
the uncertainties in the thermal (symmetric) state are multiplied by exponential factors depending on $\xi$, that is, $\Delta p_k^2 = (2 n_\mathrm{th}^k + 1) e^{2 \xi}/2$ and $\Delta x_k^2 = (2 n_\mathrm{th}^k + 1) e^{-2 \xi}/2$  
\cite{Kim89} rendering the state asymmetric (squeezed) in phase space. Here $n_\mathrm{th}^k= (e^{\beta_0 \hbar \omega_k} -1)^{-1}$ is the mean number of excitations with energy $\hbar \omega$ in a thermal reservoir at $\beta_0$. 
Consequently, state \eqref{eq:res} shows a non-zero asymmetry quantified by $\langle A_R^{(k)} \rangle_{\rho_R^{(k)}} = \hbar \omega \sinh(2 \xi) (2 n_\mathrm{th} +1)/2$. Nevertheless notice that for more general states, the operator $A_R^{(k)}$ takes into 
account both the the relative shape of the uncertainties and the relative displacements in optical phase space \cite{Man16}.


In the following, we will show that this reservoir behaves in a very similar way to a particle reservoir at thermal equilibrium, but replacing the number of particles by the asymmetry in Eq. \eqref{eq:asy}, 
and consequently the traditional chemical potential by the chemical-like potential of squeezing in Eq. \eqref{eq:chem}. This can be done by noticing how entropy behaves in the squeezed thermal reservoir.
Let us focus our discussion, without loss of generality, on a subset of modes in the reservoir. Consider an infinitesimal change in the state of this subset $\rho_R^{\prime} = \rho_R + \epsilon \Delta \rho_R$, 
where $\Delta \rho_R$ is a traceless operator accounting for the changes, and $\epsilon \ll 1$ is a real positive number. Then it follows that up to first order in $\epsilon$ we have (see Appendix \ref{appA} and Ref. \cite{QFT}):
\begin{equation} \label{eq:ent}
\Delta S_R = S(\rho_R^{\prime}) - S(\rho_R) = - \epsilon \tr[\Delta \rho_R \ln \rho_R],
\end{equation}
where $S(\rho) = -\tr[\rho \ln \rho]$ is the von Neumann entropy. Introducing Eq. \eqref{eq:gge} into \eqref{eq:ent} we arrive to the central relation:
\begin{equation} \label{eq:resent}
\Delta S_R = \beta \left( \Delta E_R - \mu \Delta A_R \right),
\end{equation}
where $\Delta E_R = \epsilon \tr[H_R \Delta \rho_R]$ and $\Delta A_R = \epsilon \tr[\sum_k A_R^{(k)} \Delta \rho_R]$ are the corresponding changes in energy and asymmetry of the reservoir.

As we will shortly see, Eq. \eqref{eq:resent} allows a physical interpretation of $\mu$ as the change in non-equilibrium free energy needed to increase the asymmetry of the reservoir by a given amount. The nonequilibrium 
free energy with respect to temperature $T$ is defined as $\mathcal{F}_R(\rho_R) \equiv \tr[H_R \rho_R] - k_B T S(\rho_R)$ \cite{Par15}. It characterizes the optimal amount of work extractable from a generic system with 
Hamiltonian $H_R$ in state $\rho_R$ with the help of a thermal reservoir at temperature $T$ \cite{Par15, Skr14, Gal16}. Identifying $\beta = 1/k_B T$ in Eq. \eqref{eq:chem} as the inverse temperature, and using 
Eq. \eqref{eq:resent}, we have that for any infinitesimal change $\rho_R \rightarrow \rho_R^{\prime}$, the change in nonequilibrium free energy is $\Delta \mathcal{F}_R = \Delta E_R - k_B T \Delta S_R = \mu \Delta A_R$. 
That is, the extractable work from this subset of modes, when access to a thermal reservoir at $\beta$ is possible, is directly proportional to the asymmetry of the subset $A_R$.
Remarkably, this leads to interpret the quantity $\mu = \Delta \mathcal{F}_R / \Delta A_R$, as the work needed to increase the asymmetry by one unit. Therefore $W_R \equiv \mu \Delta A_R$ is the analogous of a chemical work, representing 
the work needed to increase the asymmetry of the reservoir. 

Now combining Eq. \eqref{eq:resent} and the interpretation of $\mu$, we decompose the reservoir energy
\begin{equation} \label{eq:firstlaw0}
\Delta E_R = k_B T \Delta S_R + W_R \equiv Q_R + W_R,
\end{equation}
where we interpret $Q_R \equiv k_B T \Delta S_R = \Delta E_R - \mu \Delta A_R$ as the heat entering the squeezed thermal reservoir. This approach is in contrast to the usual way of identifying heat in quantum systems as the energy exchange between 
system and reservoirs \cite{Qtermo1,Qtermo2}. While the later identification is correct when handling with thermal reservoirs, the presence of further conserved quantities requires an approach based on the reservoir entropy.


\section{Thermodynamics and entropy production} \label{Sec3}
Until now we have just elucidated the properties of a squeezed thermal reservoir made up by non-interacting bosonic modes, and characterized by fixed parameters $\beta_0$ and $\xi$ (or equivalently $\beta$ and $\mu$). Now we may move to the description of the interaction between the reservoir 
and a system of interest. In particular, we consider a driven system, typically another bosonic mode with Hamiltonian $H_S (\lambda)= \hbar \omega_\lambda a_\lambda^\dagger a_\lambda$, where $\omega_\lambda$ and $a_\lambda$ may be time-dependent through the external variation of a control parameter 
$\lambda(t)$. The system interacts with the resonant modes of the reservoir through an interaction Hamiltonian $H_\mathrm{int}(\lambda)${\blue , which may also depend on $\lambda$}. 

\begin{figure}[t]
 \includegraphics[width= \linewidth]{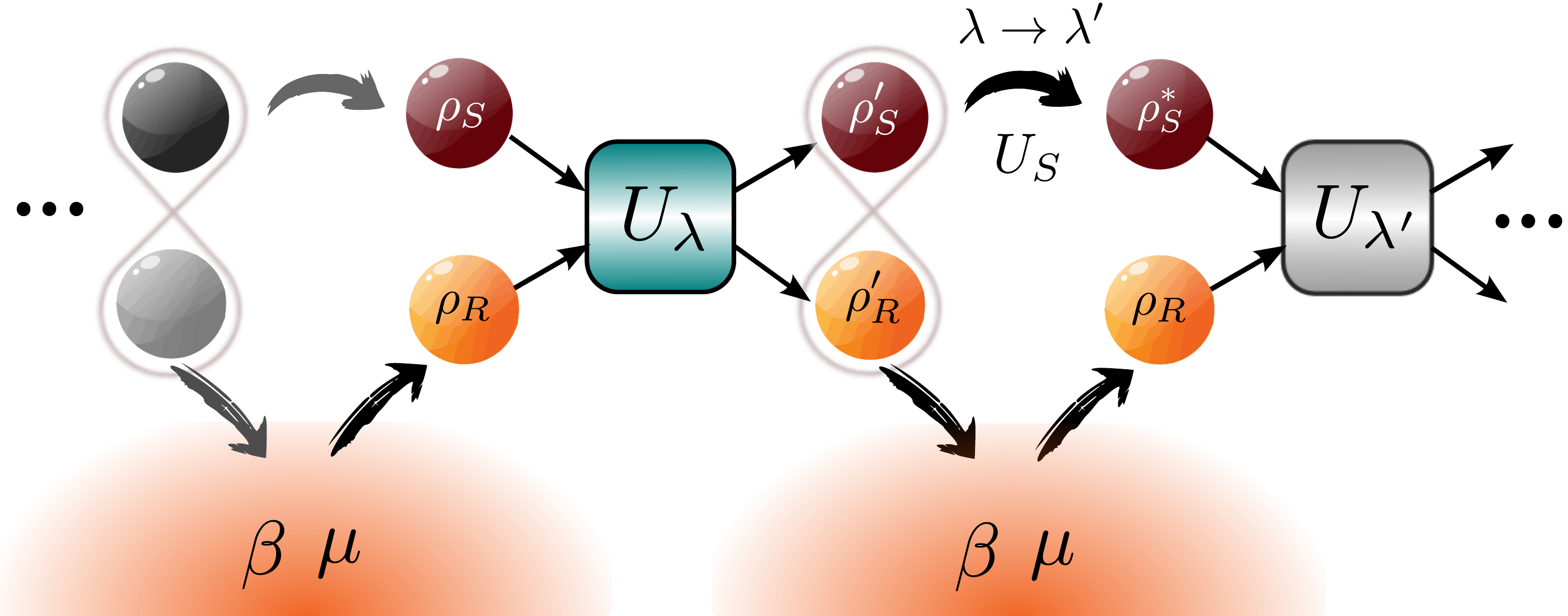}
 \caption{(color online) Sketch of the sequence of processes leading to the dynamical evolution in the main text.} \label{fig:1}
\end{figure}

In order to model the dynamical evolution we consider a sequence of infinitesimal processes, each of which can be divided in two basic steps. In the first step the system weakly interacts with resonant modes in the reservoir following a global unitary evolution $U_\lambda = \exp\big(-i \tau [H_S(\lambda) + H_R + H_\mathrm{int}(\lambda)]/\hbar \big)$.
Here we assume that this interaction occurs in a time-scale $\tau$ much faster than the external driving, so that $\lambda$ can be considered constant during the system-reservoir interaction. In addition, we notice that the much weaker interaction of the system with non-resonant modes in the reservoir can be neglected under the present circumstances \cite{Man16}. 
For any such first step in the sequence of interactions, system and reservoir start in a product state $\rho_{S} \otimes \rho_{R}$ with $\rho_{S}$ the (generic) state of the system at that time, and $\rho_R$ in Eq. \eqref{eq:gge}. The states of system and reservoir after interaction read $\rho_S^\prime = \tr_R[\rho_{SR}^\prime]$ and $\rho_R^\prime = \tr_S[\rho_{SR}^\prime]$, with $\rho_{SR}^\prime = U_\lambda (\rho_{S} \otimes \rho_{R}) U_\lambda^\dagger$. 
After that, the second step in the dynamics is performed. It consist in a smooth change of the control parameter, $\lambda \rightarrow \lambda^\prime$, producing some unitary evolution in the system alone, $U_S=\mathcal{T}_{+} \exp \big( - i {\int dt H_S[\lambda(t)]}/\hbar \big)$ with $\mathcal{T}_{+}$ the time-ordering operator. In this second step the system state change from $\rho_S^\prime$ to $\rho_S^\ast= U_S \rho_S^\prime U_S^\dagger$, while the reservoir does not 
change at all. Finally, the reservoir mode is replaced by a ``fresh'' mode in the same state $\rho_R$, and the next interaction takes place \cite{Note_reset}. 
In Fig. \ref{fig:1} we provide an schematic picture of the dynamical evolution. This kind of repeated interaction scheme can be modeled by a sequence of completely positive and trace-preserving (CPTP) maps \cite{Sca02, Pal15, Man15, Bar17} leading to the development of Lindblad master equations \cite{Sca02, Esp17, Man16} 
and quantum jump trajectories \cite{Hor12,Par13}, whose thermodynamic properties can be addressed even for arbitrary environments \cite{QFT}.

The key assumption of the dynamical evolution will be the presence of conserved quantities. The interaction between system and reservoir $H_\mathrm{int}$ should exactly preserve both energy and asymmetry 
in the global system at any time, that is $[H_S + H_R, H_\mathrm{int}] = 0$ and $[A_S + A_R, H_\mathrm{int}] = 0$. Here the asymmetry in the system $A_S$ is analogously defined as in Eq. \eqref{eq:asy} by 
$A_S(\lambda) \equiv -\hbar \omega_\lambda \left(a_\lambda^{\dagger 2} + a_\lambda^2 \right)/2$ for any value of the external parameter $\lambda$. 
The later condition for the commutators leads to the conservation of total average energy and asymmetry for the first (interaction) step of the evolution:
\begin{equation} \label{eq:conservation}
\Delta E_S + \Delta E_R = 0, ~~~~ \Delta A_S + \Delta A_R = 0,
\end{equation}
where $\Delta E_S = \tr[H_S (\rho_S^\prime - \rho_S)]$, $\Delta A_S = \tr[A_S (\rho_S^\prime - \rho_S)]$ and analogously for the reservoir. It is worth noticing that this assumption indeed prevents extra 
sources of energy or asymmetry from switching on/off interactions considered in \cite{Esp17,Bar15}.

Moreover, during the second step of the dynamical evolution energy and asymmetry may change due to the action of the external driving. 
Then on a coarse-grained time scale involving both first (interaction) and second (driving) steps, we have
\begin{align} \label{eq:non-conservation}
\dot{E}_S + \dot {E}_R = \dot{W}, ~~~~ \dot{A}_S + \dot {A}_R = \dot{\mathcal{A}}
\end{align}
where $\dot{W} = \tr[\dot{H}_S {\rho}_S]$ represents the mechanical work performed by the external driver, and analogously $\dot{\mathcal{A}} =  \tr[\dot{A}_S {\rho}_S]$ is the asymmetry induced by driving. 
Now combining Eqs. \eqref{eq:firstlaw0} and \eqref{eq:non-conservation} we state the first law of thermodynamics as
\begin{equation} \label{eq:firstlaw}
 \dot{E}_S = \dot{W} + \dot{W}_\mathrm{sq} + \dot{Q}, 
\end{equation}
where $\dot{Q} \equiv - \dot{Q}_R$ is the heat entering the system and $\dot{W}_\mathrm{sq} \equiv - \dot{W}_R = - \mu \dot{A}_R$ 
is the chemical-like squeezing work performed by the reservoir.

Conservation of energy and asymmetry in Eqs. \eqref{eq:conservation} is verified e.g. by {\blue prototypical beam splitter interactions} of the form $H_\mathrm{int} \propto a_\lambda b_k^\dagger + a_\lambda^\dagger b_k$, 
as the ones typically assumed in derivations of the master equation for the squeezed thermal reservoir \cite{Man16, WallsB,ScullyB}. 
For frozen $\lambda$, contact with the reservoir brings any initial state of the system in the long time run to the equilibrium state
\begin{equation} \label{eq:syseq}
\rho_S^\mathrm{eq}(\lambda) =  \mathcal{S}_\lambda(\xi) \frac{e^{- \beta_0 H_S}}{Z_0} \mathcal{S}_\lambda^\dagger(\xi) = \frac{e^{- \beta (H_S - \mu A_S)}}{Z_\lambda}. 
\end{equation}
Here $\mathcal{S}_\lambda(\xi) = \exp[\frac{1}{2} \xi (a_\lambda^2 - a_\lambda^{\dagger 2})]$ is the squeezing operator over the system mode, and $Z_\lambda = \tr[e^{-\beta (H_S- \mu A_S)}]$. 
More generally, including the possibility of external driving into the system [Eq. \eqref{eq:firstlaw}], the entropy production rate of a generic process as described previously can be calculated from the sum of von Neumann entropies 
of system and reservoir \cite{QFT}. That results in the following statement of the second law
\begin{align} \label{eq:ep}
\dot{S}_\mathrm{tot} &= \dot{S} + \dot{S}_R = \dot{S} + \beta (\dot{W} + \dot{W}_\mathrm{sq} - \dot{E}_S) \nonumber \\
                     &= \dot{S} + \beta (\dot{W} - \mu \dot{\mathcal{A}})  - \beta (\dot{E} _S - \mu \dot{A}_S) \geq 0,
\end{align}
{\blue where $\dot{S}$ ($\dot{S}_R$) stands for the time-derivative of the von Neumann entropy of the system (reservoir). For the derivation of Eq. \eqref{eq:ep}, we used the continuous version of Eq. \eqref{eq:resent} for the reservoir entropy changes, 
together with Eqs. \eqref{eq:non-conservation} and \eqref{eq:firstlaw}.}

\section{Reversible process} \label{Sec4}
Our objective now is to construct a generic reversible evolution for systems in contact with the squeezed thermal reservoir. We first consider the case of reversible work and asymmetry extraction by 
transforming an arbitrary initial state $\rho_S$ into the equilibrium state $\rho_S^\mathrm{eq}$ in Eq. \eqref{eq:syseq}. The transformation $\rho_S \rightarrow \rho_S^\mathrm{eq}$ might be done 
by simply letting the system relax in contact with the reservoir. However, this process is completely irreversible and prevent us from extracting any mechanical work nor asymmetry from $\rho_S$. On the contrary, 
a reversible protocol maximizing the extraction of resources can be built by extending the protocol explained in Ref. \cite{Par15} for standard thermal reservoirs (see also Ref. \cite{Man18}) to the case of a squeezed 
thermal reservoir.

This reversible process consists in two steps: 
\begin{enumerate}[label=\alph*)]
 \item An initial {\blue sudden} quench of the system Hamiltonian from $H_S(\lambda_0)$ to $H_S({\lambda_\ast})\equiv H_\ast$, where
  \begin{equation} \label{eq:protocol}
   ~~~~ H_{\ast}  =  \hbar \omega_{\lambda_\ast} a^\dagger_{\lambda_\ast} a^{~}_{\lambda_\ast} \equiv - \mathcal{S}^\dagger(\xi) \ln(\rho_S) \mathcal{S}(\xi)/\beta_0.
  \end{equation}
  This step is implemented by an instantaneous change in the control parameter $\lambda_0 \rightarrow \lambda_\ast$, which implies as well $A_S(\lambda_0) \rightarrow A_S(\lambda_\ast) \equiv A_{\ast}$. 
  The duration of this step is infinitesimal and, since the change in the control parameter is instantaneous, the system state does not change. Nevertheless, this initial quench is crucial for implementing a reversible transformation, as long as it leaves 
  the system in equilibrium with the squeezed thermal reservoir, $\rho_S =  \exp[{-\beta (H_\ast - \mu A_\ast)}]/Z_\ast$, i.e. the state $\rho_S$ is now the equilibrium state of the system for $H_\ast$ and $A_\ast$. 
  Notice that for performing this quench a work $W_\mathrm{quench} \equiv \tr[( H_{\ast} - H_S) \rho_S]$ and asymmetry $\mathcal{A}_\mathrm{quench} \equiv \tr[( A_\ast - A_S) \rho_S]$ are required, but no entropy is produced. 

 \item A reversible isothermal-like transformation where a quasi-static driving changes the Hamiltonian $H(\lambda_\ast)$ and the asymmetry operator $A (\lambda_\ast)$ back to their initial forms, while the system interacts with the reservoir. It is worth noticing that, since the system starts this step in equilibrium with the squeezed thermal reservoir, and the control parameter changes quasi-statically (infinitely slowly) from $\lambda_\ast$ to $\lambda_0$, 
 the system will be maintained in equilibrium with the reservoir at any time [Eq. \eqref{eq:syseq}], leaving the system in $\rho_S^{\mathrm{eq}}$. This implies zero entropy production during the process (reversibility), that is, $\dot{Q} = k_B T \dot{S}$ . In Appendix \ref{appB} we indicate a particular 
 realization of this quasi-static process using a concatenation of CPTP maps. 
\end{enumerate} 

In order to obtain a quantitative link between the work and asymmetry extractable from the initial state $\rho_S$, we may integrate Eq. \eqref{eq:ep} for reversible conditions (equality case) along the two steps of the protocol. For this purposes Eq. \eqref{eq:ep} can then be conveniently rewritten as  
\begin{equation} \label{eq:integration}
\beta (\dot{W}- \mu \dot{\mathcal{A}}) = - \dot{Z}_\lambda/Z_\lambda.
\end{equation}
Integrating Eq. \eqref{eq:integration} and identifying $W_\mathrm{ext} = - W_\mathrm{quench} - \int_{\lambda_\ast}^{\lambda_0} \tr[\dot{H}_\lambda \rho_\lambda^{\mathrm{eq}}]\mathrm{d}\lambda$ and $\mathcal{A}_\mathrm{ext} = - \mathcal{A}_\mathrm{quench} - \int_{\lambda_\ast}^{\lambda_0} \tr[\dot{A}_\lambda \rho_\lambda^{\mathrm{eq}}]\mathrm{d}\lambda$ 
as the total work and asymmetry extracted in the process, we obtain
\begin{align} \label{eq:landau}
 W_\mathrm{ext} - \mu \mathcal{A}_\mathrm{ext} &= {\Omega}(\rho_S) -  {\Omega}(\rho_S^{\mathrm{eq}}) \nonumber \\
                                               &= k_B T D (\rho_S || \rho_S^\mathrm{eq}) \geq 0.
\end{align}
Here the quantity $\Omega(\rho_S) \equiv \tr[(H_S - \mu A_S) \rho_S] - k_B T S(\rho_S)$ is a potential generalizing the nonequilibrium free energy for systems interacting with a squeezed thermal reservoir. 
For obtaining \eqref{eq:landau} we used that $\tr[(H_S - \mu A_S) \rho_S^\mathrm{eq}] - k_B T S(\rho_S^\mathrm{eq}) = - k_B T \ln Z_\lambda$. 
Moreover, in the last line we introduced the relative entropy $D(\rho || \sigma) = \tr[\rho (\ln \rho - \ln \sigma)] \geq 0$, which reaches zero if and only if $\rho = \sigma$ \cite{Sagawa}. 

The interpretation of Eq. \eqref{eq:landau} may be now clarified by introducing the squeezing work performed by the reservoir, $W_\mathrm{sq} = \mu (\Delta A_S + \mathcal{A}_\mathrm{ext})$. We have: 
\begin{equation}\label{eq:interpretation}
W_\mathrm{ext} - W_\mathrm{sq} +  \mu \Delta A_S =  k_B T D (\rho_S || \rho_S^\mathrm{eq}). 
\end{equation}
Eq. \eqref{eq:interpretation} tell us that any out-of-equilibrium state $\rho_S \neq \rho_S^{\mathrm{eq}}$ is a resource which may be wasted in  either extracting a positive \emph{net} amount of work, $W_\mathrm{ext} - W_\mathrm{sq}$, 
or in increasing the asymmetry of the system, $\mu \Delta A_S$. Nevertheless, notice that even when our initial state is in equilibrium, $\rho_S = \rho_S^{\mathrm{eq}}$, extraction of mechanical work is not forbidden anymore, but it is allowed by cyclic extraction 
of asymmetry from the reservoir, $W_\mathrm{ext} = W_\mathrm{sq} = \mu \mathcal{A}_\mathrm{ext}$. 

The previous protocol may be applied to more general transformations $\rho_S \rightarrow \rho_S^\prime$ by including slight modifications. In such case step a) is applied exactly as before, but 
in the quasi-static step b) we replace the final value of the control parameter by some $\lambda_\ast^\prime$ such that $H_\ast^\prime = H(\lambda_\ast^\prime) \equiv - \mathcal{S}^\dagger(\xi) \ln (\rho_S^\prime)\mathcal{S}(\xi)/\beta_0$. 
This implies that the system after quasi-static driving is now $\rho_S^\prime$. Then a final step is included: c) A final sudden quench $\lambda_\ast^\prime \rightarrow \lambda$ which instantaneously turns back the Hamiltonian and asymmetry operator 
to their original forms, $H_S$ and $A_S$. This extension is indeed equivalent to the combination of two reversible strokes, $\rho_S \rightarrow \rho_\mathrm{eq}$ as before, followed by the inversion of $\rho_S^\prime \rightarrow \rho_\mathrm{eq}$.

\section{Cyclic work extraction from a single reservoir} \label{Sec5}
Once reversible protocols have been introduced we may now discuss work extraction from a single squeezed thermal reservoir. The idea is to introduce a thermodynamic cycle on the system 
which combines unitary processes on the system and interaction with the squeezed thermal reservoir. In Ref. \cite{Man16}, we already introduced a thermodynamic cycle for work extraction: 
1) Starting with the system in $\rho_S^\mathrm{eq}$ in Eq. \eqref{eq:syseq}, the system is first detached from the reservoir and a unitary operation is implemented unsqueezing the mode $U_1 = \mathcal{S}^\dagger$. This leaves the system in the state $\rho_S = e^{- \beta_0 H_S}/Z$.
2) Then the squeezed thermal reservoir is reconnected and a simple relaxation take places, bringing the state of the system back to $\rho_S^\mathrm{eq}$. In such protocol, work is extracted in the first unitary stroke 
$W_1$, while the second stroke occurs in the absence of driving, and then the system only exchanges energy with the reservoir. Notice that this second relaxation step is intrinsically irreversible (isochoric process) and therefore it does not allow for an optimal use of the resources. 
Our aim here is to replace the isochoric process by a reversible process as introduced before. 

Let us now generalize the previous two-stroke thermodynamic cycle for work extraction. 1) As before, the first stroke consist in detaching the system from the reservoir and implementing an arbitrary initial unitary step $U_1$, leaving the system in a rather arbitrary state $\rho_S$. 
During this stroke an amount of work $W_1 = \tr[H_S (\rho_S^\mathrm{eq} - \rho_S)]$ and asymmetry $\mathcal{A}_1 = \tr[A_S (\rho_S^\mathrm{eq} - \rho_S)]$ are extracted, while the entropy production is zero.
2) Now, for the second stroke we turn back to $\rho_S^\mathrm{eq}$ in contact with the environment, but also allowing arbitrary external driving. This eventually leads to extracting some extra amount of work $W_2$ during this step, together with asymmetry $\mathcal{A}_2$. 
For this second stroke we may apply Eq. \eqref{eq:ep}. Integrating it and using that by construction $S(\rho_S) = S(\rho_S^\mathrm{eq})$, $\Delta E_S = W_1$, and $\Delta A_S = \mathcal{A}_1$, a bound for the 
total work extracted $W_\mathrm{ext} = W_1 + W_2$ is obtained
\begin{align} \label{eq:workex}
W_\mathrm{ext} \leq \mu (\mathcal{A}_1 + \mathcal{A}_2) = - \mu \Delta A_R = W_\mathrm{sq}.
\end{align}
That is, in any cyclic process one may extract an amount of work less or equal than the squeezing chemical-like work performed by the reservoir. 
This indeed provides an extra motivation to consider $W_\mathrm{sq}$ as work and the squeezed thermal reservoir as a work producing reservoir \cite{Hor16}.

The key point for reaching the equality in Eq. \eqref{eq:workex} is nothing but fully avoiding any irreversibility in the second stroke of the cycle. This is accomplished when this second stroke is exactly the (two-step) reversible process introduced below.
Applying Eq. \eqref{eq:landau} it follows that $W_2 = \Omega(\rho_S) - \Omega(\rho_S^{\mathrm{eq}}) + \mu \mathcal{A}_2 = - W_1 + \mu (\mathcal{A}_1 + \mathcal{A}_2)$, and the optimal amount of work $W_\mathrm{ext} = W_1 + W_2 = \mu (\mathcal{A}_1 + \mathcal{A}_2) = W_\mathrm{sq}$ is extracted. 
Furthermore, notice that this extraction protocol does not need any particular form of $\rho_S$ and $U_1$, e.g. we may use $U_1 = \mathbb{I}$ [cf. Eq. \eqref{eq:interpretation}]. In any case, the chemical-like squeezing work extracted from the reservoir 
equals the asymmetry extracted by the external driver. Therefore, this can be regarded as an example of a squeezing into work conversion in the context of a generalized resource framework \cite{Gur16, Los17}.

\section{Ergotropy and work extraction} \label{Sec6}
\begin{figure}[t!]
 \includegraphics[width= 1.0 \linewidth]{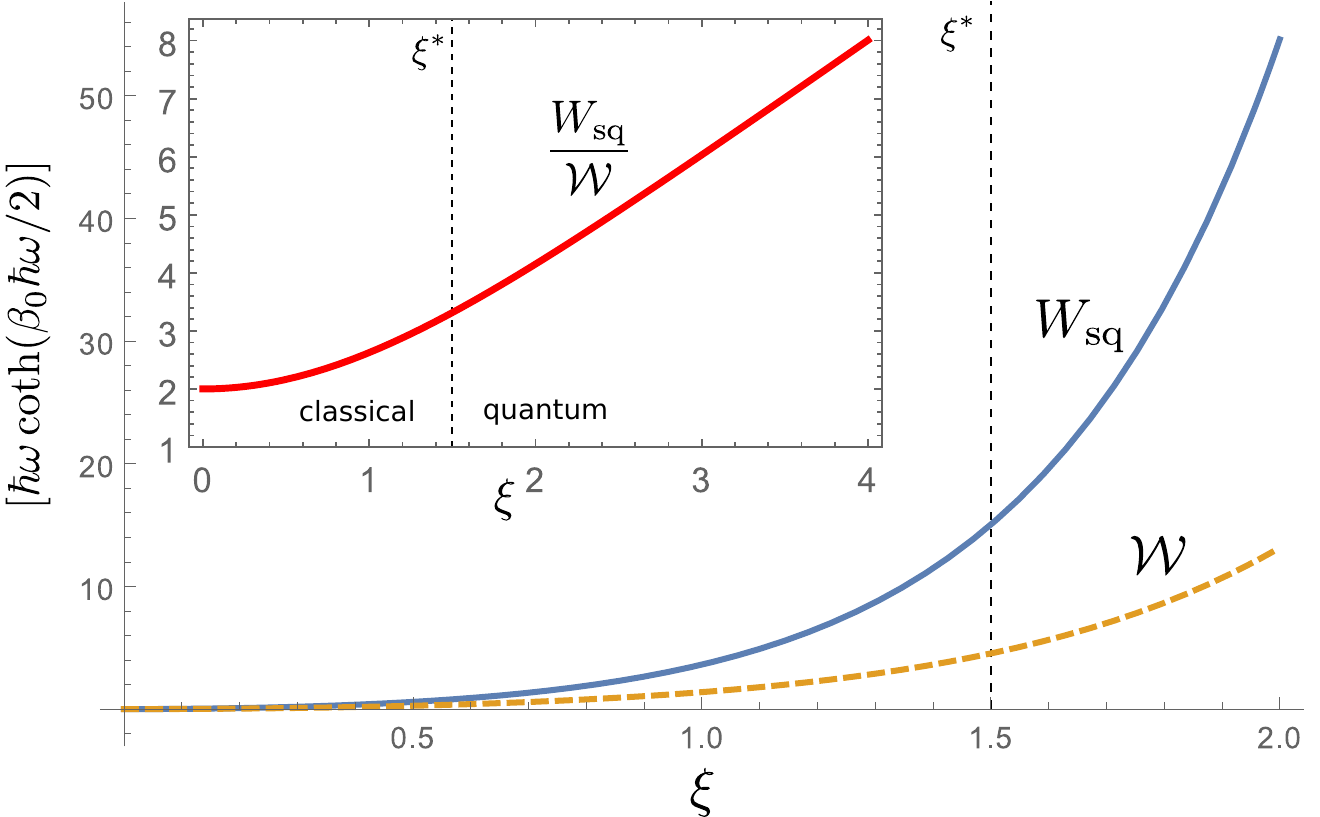}
 \caption{(color online) Comparison of the optimal work reversibly extractable from a single squeezed thermal reservoir $W_\mathrm{sq}$ (solid curve) with the protocol detailed in the text and the ergotropy $\mathcal{W}$ (dashed curve).
 Both quantities are given in units of $\hbar \omega \coth(\beta_0 \frac{\hbar \omega}{2})$. In the inset we show their ratio, which becomes more important when increasing the squeezing parameter $\xi$. In both plots the black dashed vertical line indicates 
 the boundary between classical and quantum regimes for $\hbar \omega \sim 10 k_B T_0$, leading to $\xi^\ast \sim 1.5$.}
\label{fig:2}
\end{figure}

Finally, it is worth mentioning that the maximum extractable work from a single reservoir $W_\mathrm{sq}$ is larger in general than the so-called ergotropy $\mathcal{W}$ of the non-passive state $\rho_S^{\mathrm{eq}}$ induced by the squeezed thermal 
reservoir. The later is defined as the maximum work extractable from a state using {\it only} unitary operations describing a cyclic variation of the Hamiltonian \cite{Allahverdyan}. It can be straightforwardly seen that the 
ergotropy of the state $\rho_S^{\mathrm{eq}}$ is given by $W_1$ when $U_1 = \mathcal{S}^\dagger$ \cite{Man16}. 
We notice that this is just one part of the total extractable work, $W_\mathrm{sq}= W_1 + W_2$, which can be made arbitrarily larger when increasing $\xi$ (see Fig. \ref{fig:2}).
In the following we provide explicit expressions for $\mathcal{W}$ and $W_\mathrm{sq}$.

Since $U_1 = \mathcal{S}^\dagger$, leading to $\rho_S= U_1 \rho_S^{\mathrm{eq}} U_1^\dagger = e^{-\beta_0 H_S}/Z_0$, the work extracted in the first stroke explicitly reads: 
\begin{align}
W_1 &= \tr[H_0 \rho_S^{\mathrm{eq}}] - \tr[H_0 \rho_S] \nonumber \\ 
&= \hbar \omega \sinh^2(\xi) (2 n_{\mathrm{th}} + 1) \equiv \mathcal{W}
\end{align}
where $n_{\mathrm{th}} = (e^{\beta_0 \hbar \omega} - 1)^{-1}$ is the mean number of excitations with energy $\hbar \omega$ in a thermal reservoir at $\beta_0$. Notice that $\rho_S$ is the Gibbs state, and then 
it is the state of lower energy for fix entropy. This implies that the work $W_1 = \mathcal{W}$ is by definition the ergotropy of state $\rho_S^{eq}$. Moreover, during this stroke the asymmetry extracted reads
\begin{align}
\mathcal{A}_1 &= \tr[A_0 \rho_S^{\mathrm{eq}}] - \tr[A_0 \rho_S] \nonumber \\
& = \tr[A_0 \rho_S^{\mathrm{eq}}] = \hbar \omega \sinh(2 r) (n_\mathrm{th} + 1/2),
\end{align}
where $A_0 \equiv A_S(\lambda_0)$.

Next we calculate the work and asymmetry extracted in the second (two-steps) reversible stroke of the cycle. 
Following Eq. \eqref{eq:protocol}, in the first part of the stroke (sudden quench) the Hamiltonian is modified from $H_0$ to 
\begin{equation}\label{eq:H}
H_{\ast} \equiv \hbar \omega b^\dagger b = \mathcal{S}^\dagger(\xi) H_S \mathcal{S}(\xi), 
\end{equation}
where $b = S^\dagger(\xi) a S(\xi)$, while the state of the system does not change. At this point we can see that effectively after the quench we have $\rho_S = \mathcal{S}(\xi) e^{-\beta_0 H_{\ast}}\mathcal{S}^\dagger(\xi)/Z_0$, 
which is the equilibrium state for Hamiltonian $H_{\ast}$. In this step the work extracted is 
\begin{align}
 W_2^{\mathrm{quench}} =~ & \tr[\rho_S H_0] - \tr[\rho_S H_\ast] = \tr[\rho_S H_0] \nonumber \\ 
 & - \tr[\mathcal{S}(\xi) \rho_S \mathcal{S}^\dagger(\xi) H_0] = - \mathcal{W},
\end{align}
where we used Eq. \eqref{eq:H}. This means that the ergotropy extracted in the first stroke is wasted in implementing the initial sudden quench of the second stroke. Analogously, for the extracted asymmetry we have 
$\mathcal{A}_2^{\mathrm{quench}} = - \mathcal{A}_1$, so that until now we did not gained anything at all. Nevertheless, as we will shortly see, a greater amount of work and asymmetry is going to be extracted in the quasi-static step of the stroke. 
This can be implemented by means of the slow change of a control parameter $\lambda$ from $\lambda_\ast$ to $\lambda_0$, driving the Hamiltonian back to $H_0$. 
We will parametrize this process as an unsqueezing of the Hamiltonian
\begin{equation}
 H_S(\lambda) = \hbar \omega b_\lambda^\dagger b_\lambda, ~~~~ b_\lambda = \mathcal{S}^\dagger(\lambda) a \mathcal{S}(\lambda),
\end{equation}
with $\lambda_{\ast} = \xi$ and $\lambda_{0} = 0$. These operations are usually implemented in quantum optical setups like degenerate parametric down conversion by passing the light from a pumping laser field through a photonic crystal with 
$\chi^{(2)}$ nonlinearities (see e.g. \cite{Ficek, Gerry}). This kind of driving in particular implies that
\begin{align}
 \dot{H}_S(\lambda) &= \hbar\omega (\dot{b}_\lambda^\dagger b_\lambda + b_\lambda^\dagger \dot{b}_\lambda) = -\hbar \omega \dot{\lambda} (b_\lambda^{\dagger 2} + b_\lambda^{~2}) \nonumber \\
                    &= 2 \dot{\lambda} A_S(\lambda), \\
 \dot{A}_S(\lambda) &= - \hbar \omega (\dot{b}_\lambda b_\lambda + \dot{b}_\lambda^\dagger b_\lambda^\dagger) = \hbar \omega \dot{\lambda} (2 b_\lambda^\dagger b_\lambda + 1) \nonumber \\
                    &= \dot{\lambda} (2 H_S(\lambda) + \hbar \omega),
 \end{align}
where we used $\dot{b}_\lambda = - \dot{\lambda} b_\lambda^\dagger$, following Eq. \eqref{eq:H}. In the other hand, as shown in the previous section, the quasi-static driving implies the state of the system being at any time 
$\rho_S^{\mathrm{eq}} = \mathcal{S}(\xi) e^{-\beta_0 H_S(\lambda)}\mathcal{S}^\dagger (\xi) /Z_0$. Therefore the work extracted during the quasi-static stroke reads
\begin{align}
 W_2^{\mathrm{q-s}} &= - \int_\xi^0 d\lambda \tr[A_S(\lambda) \rho_\lambda^{\mathrm{eq}}] \nonumber \\
                    &= \hbar \omega \sinh(2\xi)(2 n_\mathrm{th} + 1) \xi.
\end{align}
Analogously, for the extracted asymmetry we obtain
\begin{align}
 \mathcal{A}_2^{\mathrm{q-s}} &= - \int_\xi^0 d\lambda \tr[(2 H_S(\lambda) + \hbar \omega) \rho_\lambda^{\mathrm{eq}}] \nonumber \\ 
                              &=  \hbar \omega \cosh(2\xi) (2 n_\mathrm{th} + 1) \xi.
\end{align}

In conclusion, during a single cycle we obtain a total amount of extracted mechanical work and asymmetry 
\begin{align}
 W_{\mathrm{ext}} = \hbar \omega \xi \sinh(2\xi)(2 n_\mathrm{th} + 1) \\
 \mathcal{A}_{\mathrm{ext}} = \hbar \omega \xi \cosh(2\xi)(2 n_\mathrm{th} + 1),
\end{align}
fulfilling $W_{\mathrm{ext}} = \mu \mathcal{A}_{\mathrm{ext}}$. Notice that the squeezing work in the setup is strictly greater than the ergotropy for any value of $\xi$, 
that is $W_\mathrm{sq} \equiv \mu \mathcal{A}_{\mathrm{ext}} > \mathcal{W}$, since $\xi \sinh(2\xi) > \sinh^2(\xi) ~~\forall \xi > 0$. This can be seen in 
Fig. \ref{fig:2} where $\mathcal{W}$ and $W_\mathrm{sq}$ are plotted as a function of $\xi$, in units of $\hbar \omega \coth(\beta_0 \hbar \omega/2)$.

Finally, we point that one way of looking at the quantumness of the squeezing effects is to look at the induced uncertainty in position or momenta quadratures of our bosonic mode. In our case, for a squeezed thermal state,  
the squared uncertainty in the position quadrature $x = (1/\sqrt{2})(a + a^\dagger)$ reads
\begin{equation}
 \Delta x^2 \equiv \langle x^2 \rangle - \langle x \rangle^2 = (2 n_\mathrm{th} +1) \frac{e^{-2\xi}}{2},
\end{equation}
and the Heisenberg uncertainty principle reads $\Delta x  \Delta p \leq 1/2$. A genuine quantum state is then reached when this uncertainty falls below shot noise, i.e. when $\Delta x^2 < 1/2$.
This is equivalent to squeezing the quadrature over a certain quantity
\begin{equation}
\xi > \xi^\ast = \frac{1}{2} \ln(\coth(\beta_0 \hbar \omega/2)). 
\end{equation}
The quantity $\xi^\ast$ essentially depends on the relation between the magnitude of thermal fluctuations and the energy of the system. It diverges $\xi^\ast \rightarrow \infty$ for high temperatures $\beta_0 \rightarrow 0$, while 
it approaches zero $\xi^\ast \rightarrow 0$, when the temperature goes to zero, $\beta_0 \rightarrow \infty$. For a relevant parameter range $\beta_0 \hbar \omega \simeq [0.01, 1]$ we obtain values of $\xi^\ast$ from $2.65$ to $0.39$. 
As shown in Fig. \ref{fig:2}, larger amounts of work can be obtained in the quantum regime $\xi > \xi^\ast$, where the uncertainty in the squeezed quadrature falls below shot noise.   

\section{Conclusions} \label{Sec7}
In this paper we explored the interpretation of the squeezed thermal reservoir as a generalized equilibrium reservoir with two non-commuting charges: energy and (second-order moments) asymmetry. 
This identification can be done provided that the interaction with the reservoir preserve the two quantities, a condition easy to be fulfilled for the case of bosonic systems and in particular 
in the context of quantum optics (e.g. by a simple beam splitter interaction). Our formulation of the first and second laws of thermodynamics solve the major problem of how to correctly define work 
and heat for the squeezed thermal reservoir, but the same approach also applies to other non-thermal reservoirs such as the displaced thermal reservoir. 
In particular, our results show how previous approaches based on the naive consideration of any energy exchanged with the reservoir as heat \cite{Huang12, Lutz14, Abah14, Man16}, or in the concept of ergotropy \cite{Nie16, Nie18} fail to 
provide a complete explanation of the energetics and thermodynamics. We expect that the identification of the chemical-like squeezing work provided here together with the generic reversible protocols may have 
a strong impact in new designs of quantum thermal machines or other devices combining thermal and squeezing effects.

\begin{acknowledgments}
I would like to thank Juan M. R. Parrondo for encouraging me to do this work and Rosario Fazio for useful comments and discussions. I acknowledge financial support from the Horizon 2020 EU collaborative project QuProCS (Grant Agreement No. 641277).  
\end{acknowledgments}

\appendix

\section{Proof of Eq. (5)} \label{appA}
In the following we provide a proof of Eq.(5) in the main text, adapting the one given previously in Ref. \cite{QFT}. Consider a  change in the state of the reservoir
\begin{equation}
 \rho_R^{(k) \prime} = \rho_R^{(k)} + \epsilon \Delta \rho_R^{(k)},
\end{equation}
where $\rho_R^{(k) \prime}$ is the state of the reservoir after some arbitrary interaction with other system, $\rho_R^{(k)}$ is the original state, $\Delta \rho_R^{(k)}$ is a traceless operator 
accounting for the change, and $\epsilon \geq 0$ is a positive (small) number. In the following we will calculate the change in the von Neumann entropy associated to this process, that is, 
$\Delta S_R = S(\rho_R^{(k) \prime}) - S(\rho_R^{(k)})$.

In order to proceed we need to calculate the eigenvalues and eigenvectors of the state $\rho_R^{(k) \prime}$, which we call $\{\lambda_n^\prime, \ket{\lambda_n^\prime}\}$. This can be done by using perturbation theory when $\epsilon \ll 1$. Indeed, up to second order 
in $\epsilon$ we will have:
\begin{eqnarray} \label{eigenexpansion}
 & \lambda_n^\prime  \simeq \lambda_n + \epsilon \lambda_n^{(1)} + \epsilon^2 \lambda_n^{(2)},  \nonumber \\
 & \ket{\lambda_n^\prime} \simeq \ket{\lambda_n} + \epsilon \ket{\lambda_n^{(1)}} + \epsilon^2 \ket{\lambda_n^{(2)}},
\end{eqnarray}
where $\{ \lambda_n, \ket{\lambda_n} \}$ are the eigenvalues and eigenvectors of $\rho_R^{(k)}$, and we have also introduced the first and second order contributions. In particular the first order ones read
\begin{eqnarray} \label{eigenexpansion2}
 & \lambda_n^{(1)} = \bra{\lambda_n} \Delta \rho_R^{(k)} \ket{\lambda_n}, \\
 & \ket{\lambda_n^{(1)}} = \sum_{l \neq n} \frac{\bra{\lambda_l} \Delta \rho_R^{(k)} \ket{\lambda_n}}{\lambda_n - \lambda_l} \ket{\lambda_l}. 
\end{eqnarray}
Then we may write the change in entropy as
\begin{align}
 \Delta S_R &= S(\rho_R^{(k) \prime}) - S(\rho_R^{(k)}) \nonumber \\ 
            &= - \sum_n \lambda_n^\prime \ln \lambda_n^\prime + \sum_m \lambda_m \ln \lambda_m
\end{align}
and using Eqs. \eqref{eigenexpansion} we obtain up to second order in $\epsilon$:
\begin{equation}
 \Delta S_R \simeq - \epsilon \sum_n \lambda_n^{(1)} \ln \lambda_n - \epsilon^2 \big( \lambda_n^{(2)} \ln \lambda_n + \sum_m \frac{1}{2} \frac{\lambda_m^{(1) 2}}{\lambda_m} \big). \nonumber
\end{equation}
If we now drop the second order term and use Eq. \eqref{eigenexpansion2} we get
\begin{align}
 \Delta S_R &\simeq - \epsilon \sum_n \lambda_n^{(1)} \ln \lambda_n = - \epsilon \sum_n \bra{\lambda_n} \Delta \rho_R^{(k)} \ket{\lambda_n} \ln \lambda_n \nonumber \\ 
 &= - \epsilon \tr[\Delta \rho_R^{(k)} \ln \rho_R^{(k)}], 
\end{align}
which proofs Eq. (5) of the main text.

\section{Reversible protocol with CPTP maps} \label{appB}
Here we provide a particular realization of the protocol for reversible transformations with a squeezed thermal reservoir. It will be based on a concatenation of completely positive and trace preserving (CPTP) maps. 
As explained in the main text, for a system bosonic mode with Hamiltonian $H$ and density operator $\rho_0$, the protocol consists in a first sudden quench, where the Hamiltonian changes as $H \rightarrow H_\mathrm{eq} \equiv - \mathcal{S} \ln \rho_0 \mathcal{S}^\dagger /\beta_0$ 
while leaving the system in state $\rho$.  Then this is followed by a quasi-static process where $H_\mathrm{eq}$ transforms back to $H$. During this path the system remains in equilibrium with the squeezed thermal reservoir at any time, 
ending thus in $\rho^\mathrm{eq} = \mathcal{S} e^{- \beta_0 H} \mathcal{S}^\dagger/Z$. Notice that here for the ease of notation we drop the subscript $S$ in all quantities.

In order to give a map describing the quasi-static process, we extend the one developed in Ref. \cite{Man18}, where an isothermal processes for the case of a thermal reservoir were constructed 
by alternating infinitesimal adiabatic and isochoric steps (see also \cite{Isothermal}). Let's assume the following sequence of CPTP maps $\mathcal{E}_1 \circ \mathcal{E}_2 \circ ... \circ \mathcal{E}_N$, with $N \rightarrow \infty$. 
Each map in the sequence is intended to describe an infinitesimal time step, for which $\mathcal{E}_n(\rho_{n-1}) = \rho_n$. The changes in the Hamiltonian in each step can be also written as $H_{n-1} \rightarrow H_n$, and 
we set for consistency $H_N \equiv H$.

Now we decompose every CPTP map $\mathcal{E}$ in the sequence in the following two steps
\begin{equation} \label{eq:maps}
\mathcal{E}_n (\rho) \equiv \mathcal{G}_n \circ \mathcal{U}_n (\rho).
\end{equation}
We introduced $\mathcal{U}_n(\rho_{n-1})= \rho_{n-1}$ as a unitary sudden quench of the system Hamiltonian, where $H_{n-1} \rightarrow H_n$ instantaneously. The second step is provided by a CPTP map verifying 
$\mathcal{G}_n(\mathcal{S} \frac{e^{- \beta_0 H_n}}{Z_n}\mathcal{S}^\dagger) = \mathcal{S} \frac{e^{- \beta_0 H_n}}{Z_n} \mathcal{S}^\dagger$, that is, a generalized Gibbs-preserving map, which 
describes the interaction with the environment. During the action of $\mathcal{G}_n$ the Hamiltonian is assumed to remain constant.

The key condition to ensure a reversible process is that the change in the entropy of the system equals (minus) the change in entropy of the reservoir, here given by the last term in Eq. (13) of the main text:  
\begin{align} \label{eq:reversible}
\Delta S_n & \equiv S(\rho_n) - S(\rho_{n-1}) = \beta \tr[(H_n - \mu A_n) (\rho_n - \rho_{n-1})] \nonumber \\ 
&= \beta_0 (\tr[ \mathcal{S} H_n \mathcal{S}^\dagger (\rho_n - \rho_{n-1})]) \equiv - \beta Q_n,
\end{align}
where $A_n = \hbar \omega [p^2 - x^2]/2$ is the asymmetry. Here we have only $H_n$ (and $A_n$) in the expression because the entropy of the system only changes during the second step, $\mathcal{G}_n(\rho_{n-1}) = \rho_n$, 
when it interacts with the reservoir. This requires that the system state is close to the instantaneous equilibrium state for every step, namely $\mathcal{S} \frac{e^{-\beta H_n}}{Z_n}\mathcal{S}^\dagger$. 
In order to warranty this, we show that when assuming an infinitesimal change in the drive during any step, that is 
\begin{equation} \label{eq:hamil}
 H_n = H_{n-1} + \epsilon \Delta H_n
\end{equation}
with $\epsilon \ll 1$, then the sequence of CPTP maps defined by Eq. \eqref{eq:maps} verify Eq. \eqref{eq:reversible} up to first order in $\epsilon$.

In order to give a proof we proceed as follows. First, we will show that if the state of the systems starts close to $\rho_{n-1}^\mathrm{eq}$ before the map, then, after the application of the map $\mathcal{E}_n$ 
it remains close to the equilibrium state $\rho_{n}^\mathrm{eq}$. We can rewrite this condition as
\begin{align} \label{eq:quasistatic}
 \rho_n &= \mathcal{G}_n\left(\mathcal{S} \frac{e^{- \beta_0 H_{n-1}}}{Z_{n-1}} \mathcal{S}^\dagger + \epsilon \Delta \rho_{n-1}\right) \nonumber \\
 &= \mathcal{S}\frac{e^{-\beta_0 H_n}}{Z_n}\mathcal{S}^\dagger + \epsilon \Delta \rho_n + O(\epsilon^2),
\end{align}
where $\Delta \rho_{n-1}$ and $\Delta \rho_{n}$ are traceless operators accounting for the deviations from the equilibrium states before and after the map. If the above condition is verified, then our construction is 
self-consistent. Then, as a second step, we will proof that, since we may always rewrite $\rho_n = \rho_{n-1} + \epsilon \sigma_n$ for a suitable traceless $\sigma_n$ (in general $\Delta \rho_n \neq \sigma_n$), this implies Eq. \eqref{eq:reversible}.

We introduce Eq.  \eqref{eq:hamil} into the left-hand-side of Eq. \eqref{eq:quasistatic}. Then we use the expansions 
\begin{align}
 &e^{\epsilon \beta_0 \Delta H_n}  = 1 + \epsilon \beta_0 H_n  + O(\epsilon^2), \nonumber \\
 &Z_{n-1} = Z_n [1 + \epsilon \beta_0 \tr[\Delta H_n] + O(\epsilon^2)],
\end{align}
which combined with linearity give us the following result
\begin{widetext}
\begin{align}
& \mathcal{G}_n \left(\mathcal{S} \frac{e^{- \beta_0 H_{n-1}}}{Z_{n-1}}\mathcal{S}^\dagger + \epsilon \Delta \rho_{n-1} \right) = \mathcal{G}_n \left( \mathcal{S} \frac{e^{-\beta_0 H_n}}{Z_n} \left[ \frac{1 + \epsilon \beta_0 \Delta H_n + O(\epsilon^2)}{1 + \epsilon \beta_0 \tr[\Delta H_n] + O(\epsilon^2)}\right]  \mathcal{S}^\dagger \right)  + \epsilon ~\mathcal{G}_n(\Delta \rho_{n-1}) \nonumber \\
&  = \mathcal{G}_n \left( \mathcal{S} \frac{e^{-\beta_0 H_n}}{Z_n} \left[ 1 + \epsilon \beta_0 (\Delta H_n - \tr[\Delta H_n]) + O(\epsilon^2)\right]  \mathcal{S}^\dagger \right)  + \epsilon ~\mathcal{G}_n(\Delta \rho_{n-1}) \nonumber \\
& =  \mathcal{S} \frac{e^{-\beta_0 H_n}}{Z_n}  \mathcal{S}^\dagger + \epsilon \left[ \mathcal{G}_n(\Delta \rho_{n-1}) + \beta_0  \mathcal{G}_n(\mathcal{S} \frac{e^{-\beta_0 H_n}}{Z_n} \Delta H_n \mathcal{S}^\dagger) - \beta_0 \tr[\Delta H_n] \mathcal{S}  \frac{e^{-\beta_0 H_n}}{Z_n}\mathcal{S} ^\dagger \right] + O(\epsilon^2),
\end{align}
\end{widetext}
If we now make the identification
\begin{align} \label{identification}
 \Delta \rho_n &\equiv \mathcal{G}_n(\Delta \rho_{n-1}) + \beta_0  \mathcal{G}_n(\mathcal{S} \frac{e^{-\beta_0 H_n}}{Z_n} \Delta H_n \mathcal{S}^\dagger) \nonumber \\
 &- \beta_0 \tr[\Delta H_n] \mathcal{S}  \frac{e^{-\beta_0 H_n}}{Z_n}\mathcal{S} ^\dagger,
\end{align}
then Eq. \eqref{eq:quasistatic} is recovered. The second part of the proof now follows by obtaining the traceless matrix $\sigma_n$. Using Eq. \eqref{identification} and the above expansions it reads:
\begin{align}\label{sigman}
 \sigma_n &\equiv \rho_n - \rho_{n-1} = \epsilon \big[ \mathcal{E}_n(\Delta \rho_{n-1}) - \Delta \rho_{n-1} \\ 
 &+ \beta_0 \mathcal{E}_n( \mathcal{S} \frac{e^{-\beta_0 H_n}}{Z_n} \Delta H_n \mathcal{S}^\dagger) - \mathcal{S} \frac{e^{-\beta_0 H_n}}{Z_n}  \Delta H_n \mathcal{S}^\dagger \big], \nonumber
\end{align}
which, as expected is of order $\epsilon$. Then, we may express the eigenvalues and eigenvectors of $\rho_n$, the set $\{ p_n^k, \ket{\psi_n^k} \}$, in terms of the corresponding ones for $\rho_{n-1}$. This needs to use the relation $\rho_n = \rho_{n-1} + \epsilon~\sigma_n$, with $\sigma_n$ in Eq. \eqref{sigman}. 

We can always write
\begin{align} \label{eq:expand1}
p_n^k &= p_{n-1}^k + \epsilon \langle \psi_{n-1}^k |  \sigma_n | \psi_{n-1}^k \rangle + O(\epsilon^2), \\ \label{eq:expand12}
\ket{\psi_n}^k &= \ket{\psi_{n-1}^k} + \epsilon \sum_{l \neq k} \frac{\langle \psi_{n-1}^k |\sigma_n | \psi_{n-1}^k \rangle}{ p_{n-1}^k - p_{n-1}^l} \ket{\psi_{n-1}^l} + O(\epsilon^2).
\end{align}
On the other hand, we may obtain the same quantities to first order in $\epsilon$ from the equation $\rho_n = \mathcal{S} \frac{e^{-\beta_0 H_n}}{Z_n} \mathcal{S}^\dagger + \epsilon \Delta \rho_n + O(\epsilon^2)$. This leads to:
\begin{align}\label{eq:expand2}
p_n^k &= \frac{e^{-\beta_0 E_n^k}}{Z_n} + \epsilon \langle E_n^k | \mathcal{S}^\dagger \Delta \rho_n \mathcal{S} | E_{n}^k \rangle + O(\epsilon^2), \\ \label{eq:expand22}
\ket{\psi_n}^k &=  \mathcal{S} \ket{E_n^k} + \epsilon \sum_{l \neq k} \frac{\langle E_{n}^k | \mathcal{S}^\dagger \Delta \rho_n  \mathcal{S} | E_{n}^k \rangle}{ \frac{e^{-\beta_0 E_n^k}}{Z_n} - \frac{e^{-\beta_0 E_n^l}}{Z_n}} \mathcal{S} \ket{E_n^l} + O(\epsilon^2),
\end{align}
where $\{ E_n^k, \ket{E_n^k} \}$ are the eigenstates and eigenvectors of the Hamiltonian $H_n$. We can now calculate the change in entropy during the $n$-th step of the process, described by the map $\mathcal{E}_n$. Using Eqs. \eqref{eq:expand1} and \eqref{eq:expand12}, we obtain:
\begin{align}
\Delta S_n &= S(\rho_n) - S(\rho_{n-1}) \nonumber \\ 
           &= - \sum_k p_{n}^k \ln p_n^k + \sum_k p_{n-1}^k \ln p_{n-1}^k \nonumber \\
           &= - \epsilon \sum_k \langle \psi_{n-1}^k | \sigma_n | \psi_{n-1}^k \rangle \ln p_{n-1}^k + O(\epsilon^2).
\end{align}
Finally, by combining Eqs.\eqref{eq:expand1} and \eqref{eq:expand2}, we notice that $p_{n-1}^k = \frac{e^{-\beta_0 E_{n}^k}}{Z_n} + O(\epsilon)$. Therefore we have:
\begin{align}
\ln(p_{n-1}^k) &= \ln\big( \frac{e^{-\beta_0 E_n^k}}{Z_n}\big) + \ln[1 + O(\epsilon)] \nonumber \\ 
               &= \ln \big(\frac{e^{-\beta_0 E_n^k}}{Z_n}\big) + O(\epsilon).
\end{align}
This leads us to write
\begin{align} \label{end}
 \Delta S_n & = - \epsilon \sum_k \langle \psi_{n-1}^k |  \sigma_n | \psi_{n-1}^k \rangle \ln(\frac{e^{-\beta_0 E_n^k}}{Z_n}) + O(\epsilon^2) \nonumber \\
 &= \epsilon \beta_0 \sum_k E_n^k \langle \psi_{n-1}^k |  \sigma_n | \psi_{n-1}^k \rangle \nonumber \\
 &= - \epsilon \tr[\mathcal{S} H_n \mathcal{S}^\dagger \sigma_n] + O(\epsilon^2),
\end{align}
where, in the last step, we have used $\ket{p_{n-1}^k} = \ket{p_{n}^k} +  O(\epsilon) = \mathcal{S} \ket{E_n^k} + O(\epsilon)$, which follows from Eqs. \eqref{eq:expand12}
and \eqref{eq:expand22} and the cyclic property of the trace. Eq. \eqref{end} corresponds to Eq. \eqref{eq:reversible} up to first order in $\epsilon$, and completes the proof.


\begin{thebibliography}{10}
\bibitem{Note_reset} Notice that since this ``fresh'' mode is freely provided by the reservoir, the replacement does not incurs in any extra energetic cost. Moreover, since the state of the whole 
reservoir changes only infinitesimally due to the interaction with the system, no extra entropy is produced \cite{QFT}.
\end{thebibliography}
\end{document}